\documentclass[twocolumn,showpacs,preprintnumbers,amsmath,amssymb,APSl,prd,nofootinbib,superscriptaddress]{revtex4-2}
\usepackage{graphicx,color}
\usepackage{subfig}
\usepackage{amssymb}
\usepackage{hyperref}
\usepackage[utf8]{inputenc}
\usepackage[english]{babel}
\usepackage{epsfig}
\usepackage{wasysym}
\usepackage{color,xcolor}
\usepackage{amsmath}
\usepackage{bm}
\usepackage{epsfig}
\usepackage{amsfonts}
\usepackage{dcolumn}
\usepackage{float}

\hypersetup{colorlinks=true, linkcolor=blue, citecolor=green}

\begin{document}

\title{Spontaneous Lorentz symmetry-breaking constraints in Kalb-Ramond gravity}

	\author{Ednaldo L. B. Junior} \email{ednaldobarrosjr@gmail.com}
\affiliation{Faculdade de F\'{i}sica, Universidade Federal do Pará, Campus Universitário de Tucuruí, CEP: 68464-000, Tucuruí, Pará, Brazil}

     \author{José Tarciso S. S. Junior}
    \email{tarcisojunior17@gmail.com}
\affiliation{Faculdade de F\'{i}sica, Programa de P\'{o}s-Gradua\c{c}\~{a}o em F\'{i}sica, Universidade Federal do Par\'{a}, 66075-110, Bel\'{e}m, Par\'{a}, Brazill}

	\author{Francisco S. N. Lobo} \email{fslobo@ciencias.ulisboa.pt}
\affiliation{Instituto de Astrof\'{i}sica e Ci\^{e}ncias do Espa\c{c}o, Faculdade de Ci\^{e}ncias da Universidade de Lisboa, Edifício C8, Campo Grande, P-1749-016 Lisbon, Portugal}
\affiliation{Departamento de F\'{i}sica, Faculdade de Ci\^{e}ncias da Universidade de Lisboa, Edif\'{i}cio C8, Campo Grande, P-1749-016 Lisbon, Portugal}

    \author{\\Manuel E. Rodrigues} \email{esialg@gmail.com}
\affiliation{Faculdade de F\'{i}sica, Programa de P\'{o}s-Gradua\c{c}\~{a}o em F\'{i}sica, Universidade Federal do Par\'{a}, 66075-110, Bel\'{e}m, Par\'{a}, Brazill}
\affiliation{Faculdade de Ci\^{e}ncias Exatas e Tecnologia, Universidade Federal do Par\'{a}, Campus Universit\'{a}rio de Abaetetuba, 68440-000, Abaetetuba, Par\'{a}, Brazil}

 \author{Diego Rubiera-Garcia} \email{ drubiera@ucm.es}
\affiliation{Departamento de Física Téorica and IPARCOS, Universidad Complutense de Madrid, E-28040 Madrid, Spain}

     \author{Luís F. Dias da Silva} 
        \email{fc53497@alunos.fc.ul.pt}
\affiliation{Instituto de Astrof\'{i}sica e Ci\^{e}ncias do Espa\c{c}o, Faculdade de Ci\^{e}ncias da Universidade de Lisboa, Edifício C8, Campo Grande, P-1749-016 Lisbon, Portugal}

    \author{Henrique A. Vieira} \email{henriquefisica2017@gmail.com}
\affiliation{Faculdade de F\'{i}sica, Programa de P\'{o}s-Gradua\c{c}\~{a}o em F\'{i}sica, Universidade Federal do Par\'{a}, 66075-110, Bel\'{e}m, Par\'{a}, Brazill}

\begin{abstract}

In this work, we study timelike and lightlike geodesics in Kalb-Ramond (KR) gravity around a black hole with the goal of constraining the Lorentz symmetry-breaking parameter $l$. The analysis involves studying the precession of the S2 star periastron orbiting Sgr A* and geodesic precession around the Earth. The ratio of precession frequencies for General Relativity (GR) and KR gravity is computed, with Event Horizon Telescope (EHT) results providing a parameter range for the spontaneous symmetry-breaking of $-0.185022 \leq l \leq 0.060938$. Utilizing the geodesic precession frequency from the Gravity Probe B (GP-B), the $l$ parameter is further constrained to $-6.30714 \times 10^{-12} \leq l \leq 3.90708 \times 10^{-12}$, which is consistent with the Schwarzschild limits. Moreover, for timelike geodesics, the innermost circular orbit (ICO) and innermost stable circular orbit (ISCO) are determined and analyzed to illustrate the impact of the symmetry breaking term. Zoom-whirl obstructions are compared with the Schwarzschild solution.  Lower and upper limits of the photon sphere for lightlike geodesics are established to demonstrate the influence of KR gravity on the photon sphere.  Additionally, the shadow radius is determined for two observers, one situated at a finite distance from the KR black hole, and the other located at an infinite distance, to constrain the symmetry-breaking parameter $l$, with comparisons made to EHT results.  The bounds for $l$ derived from constraints on the photon sphere radius for lightlike geodesics yield $-0.0700225 \leq l \leq 0.189785$ using EHT data.
The findings of this paper align with experimental results in the $l \rightarrow 0$ limit.
\end{abstract}

\date{\today}

\maketitle

\section{Introduction}

Einstein's General Theory of Relativity (GR) has obtained substantial experimental support since its initial confirmation in 1919. This initial validation stemmed from observations during the total solar eclipse of May 29, 1919, wherein the gravitational deflection of starlight near the Sun was measured by two separate expeditions in the West African island of Príncipe \cite{Dyson:1920cwa}, and the Brazilian town of Sobral \cite{Eclipse, Crispino:2020txj}.
In 2015, the Laser Interferometer Gravitational-Wave Observatory (LIGO) collaboration confirmed Einstein's prediction of gravitational waves resulting from the collision of two black holes located 1.3 billion light years from Earth \cite{LIGO,LIGO2}.
Recent confirmations of GR include the discovery of the black hole image at the center of the galaxy M87 in 2019 \cite{EHT1,EHT2,EHT3,EHT4,EHT5, EHT6} and the image of the black hole at the center of our galaxy, Sagittarius A* (Sgr A*), both achieved through the efforts of the Event Horizon Telescope (EHT) collaboration \cite{EHT7,EHT8,EHT9,EHT10,EHT11,EHT12,EHT13,EHT14,EHT15}. 
Furthermore, the precession of the star S2, which orbits very close to the event horizon of Sgr A*, was observed for the first time in 2018 by the GRAVITY collaboration \cite{GRAVITY1,GRAVITY4,GRAVITY5}, aligning with predictions made by GR and the Schwarzschild metric.

The detection of stars surrounding Sgr A* and recent observations have paved the way for the study of precession and periodic motions around black holes. In \cite{P1}, the investigation focuses on periodic geodesics of precession and periodic orbits of time-like particles within a black-bounce scenario, demonstrating how variations in the bounce parameter can alter these orbits. Similarly, \cite{P3} examines the case of a Reissner-Nordström black-bounce, deriving marginally bound and innermost stable circular orbits. This study revealed that minor changes in charge and the bounce parameter can induce transitions between periodic orbits and precession.  A related analysis in \cite{P2} delved into the behavior of a hairy black hole in Horndeski theory, establishing constraints using precession data from the S2 star obtained by GRAVITY. It demonstrated that slight alterations in the coupling parameter can cause neutral particles to oscillate between quasi-periodic orbits, periodic orbits, or no orbits at all. 

Furthermore, in \cite{EHT}, it was demonstrated that images of Sgr A* obtained by the EHT impose stringent constraints on model parameters predicting a shadow size larger than that of a Schwarzschild black hole at a given mass. 
In 2004, the Gravity Probe B (GP-B) experiment aimed at testing two key predictions of GR, namely, geodetic effects and structural drag, using cryogenic gyroscopes in Earth orbits \cite{GPROBB}. Results revealed a geodesic drift rate of $-6601.8\pm 18.3$mas/yr, consistent with GR's predicted value of $-6606.1$mas/yr, and a frame drag rate of $-37.2\pm 7.2$mas/yr, consistent with the predicted $-39.2$mas/yr by GR \cite{GPB}. Recently, in \cite{presgeo}, it was shown that GP-B results can constrain parameters of the solution for static, spherically symmetric black holes.

Despite these experimental success, GR comes with its own inconsistencies, most notably its incompatibility with quantum mechanics, the unavoidable presence of space-time singularities in some of the most iconic solutions (such as black holes or early cosmological evolution), or the information loss problem in black holes. Attempts to quantize gravity require an understanding of Planck-scale phenomena, a domain explored by string theories \cite{corda1, corda2, corda3, corda4} and loop quantum gravity \cite{LQG1,LQG2,LQG3,LQG4}. These theories, alongside others like noncommutative field theory \cite{NC,NC1,NC2}, massive gravity \cite{MG1,MG2}, $f(T)$ gravity \cite{fT1,fT2, fT3, fT4, fT5, fT6, fT7, fT8, fT9, fT10, fT11, fT12}, and Horava-Lifshitz gravity \cite{Horava}, incorporate the breaking of Lorentz symmetry into their formulations. Lorentz symmetry, a foundational principle of GR, links all local physical references. Its violation can manifest in two ways: (i) explicit violation, where the theory's Lagrangian lacks invariance under Lorentz transformations,  where the theories exhibit incompatibility with the Bianchi identity \cite{Bianch}; (ii) and spontaneous violation, where the Lagrangian remains invariant, but physical phenomena exhibit symmetry breaking \cite{Lorentz1,Lorentz2, Lorentz3, Lorentz4, Lorentz5, Lorentz6, Lorentz7, bumblebee,Assuncao:2019azw}.

In the context of gravitational theories, the spontaneous breaking of Lorentz symmetry was addressed in \cite{bumblebee1}, assuming couplings between Riemannian spacetime and the bumblebee field, responsible for the symmetry breaking. This approach aimed to investigate gravitational phenomena such as perihelion advance, light deflection, and Shapiro time. Keer's exact solution within this framework was presented in \cite{bumblebee4}. The shadow of a black hole with a cosmological constant in Bumblebee gravity was determined in \cite{bumblebee3}, and for a Keer-Sen type black hole in \cite{bumblebee2}. Additionally, a wormhole solution for the Bumblebee gravity model was derived in \cite{bumblebee5}. 

In bosonic string theories, particular attention is given to the Kalb-Ramond (KR) field, a second-order tensor field \cite{KRoriginal}. This field spontaneously breaks Lorentz symmetry through non-minimal coupling with the Ricci scalar, resulting in a non-zero vacuum expectation value \cite{VLorentz, VLorentz2, VLorentz3}. Over the years, the KR field has been investigated within the gravitational framework, coupled with GR \cite{KRteste}. Such studies aimed to assess observable signatures, including the effects of gravitational lensing and the precession of the perihelion. Furthermore, the possibility of inflation with antisymmetric tensor field KR models coupled to gravity was investigated in \cite{KRinfla}. In the context of strong lenses for extra dimensions, the KR field was explored \cite{KRlenteforte}, and rotation effects were accounted for in subsequent studies \cite{KRKumar}.

Additionally, the KR parameter's influence on the motion of both massive and non-massive particles around black holes was examined, along with the determination of the horizon structure, photon orbit characteristics, innermost stable circular orbit (ISCO), shadow size, weak lensing effects, and the effective potential of an effect constructed by analogy with the electromagnetic field \cite{KRparticulas}. Recently, the violation of gravitational parity due to the coupling of the dual Riemann curvature to the 2-form KR field was investigated \cite{KRparidade}, along with its potential role as a candidate for dark matter \cite{KRdarkmatter}.

In \cite{KR}, the authors derived static and spherically symmetric solutions for Schwarzschild-like configurations, both with and without a cosmological constant, incorporating the KR field. These solutions demonstrate the spontaneous breaking of Lorentz symmetry in gravity, as the KR field attains a non-zero vacuum expectation value.
Their approach features a self-interacting KR field in the non-minimally coupled Einstein-Hilbert action  \cite{KRacoplado1, KRMaluf}. The action of the theory is given by
\begin{eqnarray}
 S=\int d^4x\sqrt{-g}\Bigg[ R-\frac{1}{6}H^{\mu\nu\rho}H_{\mu\nu\rho}-V(B^{\mu\nu}B_{\mu\nu})\nonumber\\
 +\zeta_2 B^{\rho\mu}B^{\nu}_{\;\;\mu}R_{\rho\nu}+\zeta_3 B^{\mu\nu}B_{\mu\nu}R\Big]\,.    
\end{eqnarray}
Here $H_{\mu\nu\rho}=\partial_{[\mu}B_{\nu\rho]}$ is the strength of the KR field, and $B_{\mu\nu}$ is the KR field. If we consider the expectation value of the field contraction in vacuum as $\langle B_{\mu\nu}\rangle =b_{\mu\nu}, \langle B^{\mu\nu}B_{\mu\nu}\rangle =\mp b^2$, the coupling term $\zeta_3$ in the action is reduced to a linear term in $R$, which can be converted into an Einstein-Hilbert term by a coordinate transformation. We will therefore not consider it here. If we now vary the action with respect to the metric, we obtain
\begin{eqnarray}
&& R_{\mu\nu}-\frac{1}{2}g_{\mu\nu}R=\frac{1}{2}H_{\mu\alpha\beta}H_{\nu}^{\;\;\alpha\beta}-\frac{1}{12}g_{\mu\nu}H^{\alpha\beta\rho}H_{\alpha\beta\rho}\nonumber\\
&& +2V'(X)B_{\alpha\mu}B^{\alpha}_{\;\;\nu}-g_{\mu\nu}V(X)+\zeta_2\Bigg[\frac{1}{2}g_{\mu\nu}B^{\alpha\rho}B^{\beta}_{\;\;\rho}R_{\alpha\beta}\nonumber\\
&& -B^{\alpha}_{\;\;\mu}B^{\beta}_{\;\;\nu}R_{\alpha\beta}-B^{\alpha\beta}B_{\nu\beta}R_{\mu\alpha}-B^{\alpha\beta}B_{\mu\beta}R_{\nu\alpha}\nonumber\\
&&  +\frac{1}{2}\nabla_{\alpha}\nabla_{\mu}\left(B^{\alpha\beta}B_{\nu\beta}\right)+\frac{1}{2}\nabla_{\alpha}\nabla_{\nu}\left(B^{\alpha\beta}B_{\mu\beta}\right)\nonumber\\
&& -\frac{1}{2}\nabla^{\alpha}\nabla_{\alpha}\left(B_{\mu}^{\;\;\rho}B_{\nu\rho}\right)-\frac{1}{2}g_{\mu\nu}\nabla_{\alpha}\nabla_{\beta}\left(B^{\alpha\rho}B^{\beta}_{\;\;\rho}\right)\Bigg ]    \,.
\end{eqnarray}
Decomposing the KR field into $B_{\mu\nu}=\hat{E}_{[\mu}v_{\nu]}+\epsilon_{\mu\nu\alpha\beta}v^{\alpha}\hat{B}^{\beta}$, where $v^{\alpha}$ is a time-like quadri-vector, and $\hat{E}^{\mu}$ and $\hat{B}^{\mu}$ are the space-like electric and magnetic pseudo-fields. We will assume spherical and static symmetry, so the metric is given by 
\begin{eqnarray}
ds^2=-A(r)dt^2+B(r)dr^2+r^2\left(d\theta^2+\sin^2\theta d\phi^2\right).    
\end{eqnarray}
For a solution with only $\hat{E}\equiv \hat{E}(r)=b_{10}$, we have the 2-form $b_{(2)}=-\hat{E}(r)dt\wedge dr$. In this particular case, we have $H_{\mu\nu\alpha}\equiv 0$, or $H_{(3)}=db_{(2)}=0$. Choose $V(x)=\lambda X^2/2$, where $X=B^{\mu\nu}B_{\mu\nu}+b^2$, so $V'\equiv 0$. We then have $\hat{E}(r)=|b|\sqrt{A(r)B(r)/2}$ and $b^{\mu\nu}b_{\mu\nu}=-b^2$, so the equations of motion are given by
\begin{eqnarray}
&& 2\frac{A''}{A} -\frac{A'B'}{AB}-\frac{A'^2}{A^2} +\frac{4A'}{rA}=0\\
&& 2\frac{A''}{A} -\frac{A'B'}{AB}-\frac{A'^2}{A^2} -\frac{4B'}{rB}=0\\
&&2\frac{A''}{A} -\frac{A'B'}{AB}-\frac{A'^2}{A^2} +\frac{(1+l)}{lr}\left(\frac{A'}{A}-\frac{B'}{B}\right)\nonumber\\
&& -2\frac{B}{lr^2}+\frac{2(1-l)}{lr^2}=0\label{eq1}\,,
\end{eqnarray}
where we define $l=\zeta_2 b^2/2$. Subtracting the first equation from the second one we get
\begin{eqnarray}
\frac{A'}{A}+\frac{B'}{B}=0\,,    
\end{eqnarray}
whose solution is
\begin{eqnarray}
 B(r)=\frac{c_1}{A(r)}\,.   
\end{eqnarray}
Substituting this solution into \eqref{eq1}, and integrating we have
\begin{eqnarray}
    A(r)=\frac{c_1}{1-l}+\frac{c_2}{r}\,.
\end{eqnarray}
The constant $c_2$ can be calculated from the Komar mass, which leads to $c_2=-2M$. To revert to the Schwarzschild solution, if we have the coupling with the zero KR field, $l=0$, we must then have $c_1=1$. In this case, the solution is as follows
\begin{eqnarray}
ds^2 &= & -\left(\frac{1}{1-l}-\frac{2M}{r}\right)dt^2+\left(\frac{1}{1-l}-\frac{2M}{r}\right)^{-1}dr^2
	\nonumber \\
&&+ r^2(d\theta^2+\sin^2\theta d\varphi^2)\,,\label{KR0}
\end{eqnarray}
where $l$ is an additional parameter that characterizes the spontaneous breaking of Lorentz symmetry.
This solution was first obtained from the reference \cite{KR}. Note that the new parameter $l=\zeta_2 b^2/2$, which is the product of the coupling constant and the square of the expectation value of the KR field, can in principle take any real value. To keep the signature of the  metric $(-,+,+,+)$ out of the event horizon, we start with the single boundary condition $0\leq l< 1$. Since the solution is asymptotically flat, the closer the value of $l$ is to unity, the further the solution deviates from the Minkowski solution at infinity, demonstrating the breaking of Lorentz symmetry.

Additionally, the impact of the solutions on thermodynamic properties and the implications of Lorentz violation were investigated through classical gravitational experiments within the solar system. These experiments provided constraints on the Lorentz violation parameter. 
Here, we will consider the line element \eqref{KR0} to denote a massive object and impose constraints to $l$.

In this work, we explore timelike and lightlike geodesics for a black hole solution in KR gravity. The aim is to constrain the Lorentz symmetry-breaking parameter by analyzing the periastron precession of the S2 star  orbiting Sgr A*. The geodesic precession around Earth will also be considered for this purpose. Note that timelike geodesics lead to the determination and analysis of the innermost circular orbit (ICO) and the innermost stable circular orbit (ISCO), revealing the impact of the symmetry breaking term. The lightlike geodesics help to identify the lower and upper limits of the photon sphere, highlighting KR gravity's influence. Finally, we determine the shadow radius for two observers to constrain the symmetry breaking parameter and compare with the results from the EHT.

This paper is organised in the following manner: In Sec. \ref{sec:two}, the general formalism for the precession of timelike geodesics is presented. The ratio of precession frequencies for GR and KR gravity is calculated, and the results of the EHT for the precession of the star S2 around Sgr A* are analyzed, yielding a parameter range for $l$. Section \ref{Geodesic Precession} uses the geodesic precession frequency of a gyroscope from the Gravity Probe B (GP-B) to constrain the $l$ parameter of the KR solution. 
Section \ref{time-like} investigates periodic ICOs, ISCO, and marginally connected orbits for timelike geodesics, demonstrating their variation with respect to the symmetry breaking parameter. 
Section \ref{LIGHT-LIKE} presents bounds for $l$ derived from constraints on the radius of the photon sphere for lightlike geodesics.  Finally, in Sec. \ref{Sec:Conclusion}, we summarize our work and conclude.
Geometrized units ($G=1$ and $c=1$) and the metric signature ($-,+,+,+$) are assumed throughout.

\section{Orbital precession}\label{sec:two}

\subsection{General formalism for precession}

In GR, the motion of a free particle is governed by the Lagrangian $\mathcal{L}=\frac{1}{2}g_{\mu\nu}\dot{x}^{\mu}\dot{x}^{\nu}$, where the overdot denotes differentiation with respect to an affine parameter. When a test particle moves along timelike geodesics, it must satisfy the condition $\mathcal{L}=-1$, and therefore we have \cite{Inverno}
\begin{equation}
g_{\mu\nu}\dot{x}^{\mu}\dot{x}^{\nu}=-1.
\label{gmn}
\end{equation}

A general static and spherically symmetric metric can be written as
\begin{equation}
    ds^2 = -A(r) dt^2 +B(r)dr^2 + C(r)d\theta^2 + D(r)\sin^2 \theta d\varphi^2\,.
    \label{eq:dsgeral}
\end{equation}
Without loss of generality, consider that a specific geodesic lies in the equatorial plane $\theta= \pi /2$, with $\dot{\theta}=0$, and the specific configuration is provided by
\begin{equation}
    -A(r)\dot{t}^2+B(r)\dot{r}^2+D(r)\dot{\varphi}^2=-1\,.
    \label{eq:dsmov}
\end{equation}
When we multiply Eq. \eqref{eq:dsmov} by $A(r)$ and take into account $D(r)\dot{\varphi}=L$ and $A(r)\dot{t}=E$, where $L$ and $E$ denote the angular momentum and the energy of the particle, respectively, derived from the Euler-Lagrange equation, we obtain:
\begin{equation}
\dot{r}^2=\frac{1}{A(r)B(r)}\left[E^2-A(r)\left(1+\frac{L^2}{D(r)}\right)\right]\,,\label{rponto}
\end{equation}
which can be rewritten as
\begin{equation}
\frac{d\varphi}{dr}=\frac{\sqrt{A(r)B(r)}}{D(r)}L\left[E^2-A(r)\left(1+\frac{L^2}{D(r)}\right)\right]^{-1/2}\,.\label{dvarphi}
\end{equation}

To derive the relativistic orbital periastron precession, we compare it with the GR result for the Schwarzschild metric of the precession of the star S2 around Sgr A* obtained by the GRAVITY collaboration \cite{GRAVITY1}. To do this, we calculate the orbital precession $\delta\omega$ by \cite{P1,P2,  P3}
\begin{equation}
\delta\omega=\delta\phi-2\pi \,, 
\label{deltaomega}
\end{equation} 
where $\delta\phi$ is the total angle accumulated in one orbital period with respect to the equatorial plane, given by
\begin{equation}
\delta\phi=2\int^{r_a}_{r_p}\frac{d\varphi}{dr}dr=2\int^{\pi}_{0}\frac{d\varphi}{d\chi}d\chi \,. 
\label{intdelta}
\end{equation} 
Here, $r_a=a(1+e)$ and $r_p=a(1-e)$ are the apoastron and periastron depending on the semi-major axis $a$ and the eccentricity $e$ of the orbits. This results from the parameterization of the coordinate $r$ depending on the relativistic anomaly $\chi$, so that
\begin{equation}
r=\frac{a(1-e^2)}{1+e\cos\chi}\,,\label{cord1}
\end{equation}
where $r_p$ corresponds to $\chi=0$ and $r_a$ to $\chi=\pi$, respectively.

Using $\frac{d\varphi}{d\chi}=\frac{d\varphi}{dr}\frac{dr}{d\chi}$ and Eq. \eqref{cord1}, we rewrite Eq. \eqref{dvarphi} as:
\begin{eqnarray}
&&\frac{d\varphi}{d\chi}=\frac{\sqrt{A(\chi)B(\chi)}}{D(\chi)}a\,e\, L\frac{\left(1-e^2\right)\sin\chi}{\left(1+e\cos\chi\right)^2}\nonumber\\
&&\times\left[E^2-A(\chi)\left(1+\frac{L^2}{D(\chi)}\right)\right]^{-1/2}\,. \label{2d}
\end{eqnarray}
Since the radial velocity at the return points is zero, i.e. $\dot{r}^2\big|_{r_{a,p}}=0$, the angular momentum and the energy can be respectively written as follows:
\begin{equation}
L^2=\frac{D(r_a)D(r_p)\left[A(r_p)-A(r_a)\right]}{D(r_p)A(r_a)-D(r_a)A(r_p)}\,,
\label{L}
\end{equation}
and
\begin{equation}
E^2=\frac{A(r_p)A(r_a)\left[D(r_p)-D(r_a)\right]}{A(r_a)D(r_p)-A(r_p)D(r_a)}\,.
\label{E}
\end{equation}

\subsection{Ratio between frequencies}

In order to obtain $\delta\omega$, we assume that Sgr A* is a black hole and the spherically symmetric solution of KR gravity \cite{KR}, given by
\begin{equation}
A(r)=\frac{1}{B(r)}=\left(\frac{1}{1-l}-\frac{2M}{r}\right)\,, \quad 
 C(r)=D(r)=r^2\,.
\label{dsKB}
\end{equation}
If we substitute Eqs. \eqref{L} and \eqref{E} into Eq. \eqref{2d}, consider \eqref{cord1} and integrate by \eqref{intdelta}, we get Eq. \eqref{deltaomega}, in which we treat the effects of $l$ as a pertubation, we arrive at the following orbital precession: 
\begin{equation}
\delta\omega_{KB}=2\pi\sqrt{1-l}+\frac{6M\pi(1-l)^{3/2}}{a(1-e^2)}-2\pi\,.
\label{omegaKB}
\end{equation}
Here we immediately realize that for the limit $l\rightarrow 0$, we get the result predicted by GR, i.e., 
\begin{equation}
\delta\omega_{GR}=\frac{6M\pi}{a(1-e^2)}\,.\label{omegaSC}
\end{equation}

To constrain the Lorentz symmetry-breaking parameter $l$, we take the ratio between the orbital precessions \eqref{omegaKB} and \eqref{omegaSC}, such that
\begin{equation}
f=\frac{\delta\omega_{KB}}{\delta\omega_{GR}}=(1-l)^{3/2}+(1-e^2)\frac{a\left(\sqrt{1-l}-1\right)}{3M}\,,\label{f}
\end{equation}
where in the limit $l \rightarrow 0$, we have $f \rightarrow 1$, as expected. Let us now assume values determined by the GRAVITY collaboration \citep{GRAVITY1} for the orbital precession of the star $S2$ around Sgr A*.  The parameters $e=0.885$ and $a=1031$ are taken into account, as well as the mass $M_{Keck}= 3.951 \times 10^6$M$_\odot$ \cite{MASSA}, for which the values for $f$ provided by \cite{GRAVITY1} are $f=1.10 \pm 0.19$, which we can compare with \eqref{f}, giving us a range of values for the parameter $l$ such that
\begin{eqnarray}
-0.185022\leq l\leq 0.060938\,. 
\end{eqnarray}
This result for $l$-values is 10 to 11 orders of magnitude larger than that obtained by \cite{KR} for the Mercury perihelion.

\section{Geodesic Precession}\label{Geodesic Precession}

 In this section, we compare the results of geodetic precession obtained by Gravity Probe B (GP-B)\cite{GPB} in a polar orbit at an altitude of 642 km around the Earth. 

Assume that a test particle moves along a timelike geodesic with 4-velocity $\dot{x}^{\mu}=dx^{\mu}/d\tau$. From Eq. \eqref{rponto}, using the symmetry $A(r)=B(r)^{-1}$ and $D(r)=r^2$ we obtain
\begin{eqnarray}
\dot{r}^2+A(r)\left(1+\frac{L^2}{r^2}\right)=E^2\,,
\end{eqnarray}
where we identify the effective potential as
\begin{eqnarray}
V_{\rm eff}=A(r)\left(1+\frac{L^2}{r^2}\right).\,\label{Veff}
\end{eqnarray}

For a stable circular orbit, we have $\dot{r}=0$ and $dV_{\rm eff}/dr=0$, which implies that the 4-velocities $\dot{r}$ and $\dot{\varphi}$ are determined by:
\begin{eqnarray}
\dot{r} &=& \sqrt{\frac{2}{2A(r)-A'(r)r}}\,,\\
\dot{\varphi} &=& \sqrt{\frac{A'(r)}{2rA(r)-A'(r)r^2}}\,,
\end{eqnarray}
respectively.

Now, we define the orbital angular velocity of the test particle as:
\begin{equation}
\Omega\equiv\frac{\dot{\varphi}}{\dot{r}}=\sqrt{\frac{A'(r)}{2r}}\,.\label{vangular}
\end{equation}
The test particle's proper angular velocity is defined by $L=r^2\dot{\varphi}=r^2\omega$, and so we have that 
\begin{eqnarray}
\omega=\Omega\left[A(r)-\frac{r}{2}A'(r)\right]^{-1/2}\,.\label{angularpropria}
\end{eqnarray} 

To determine the geodetic precession frequency $\Theta_{\rm GP}$ for a gyroscope within the GP-B probe, we utilize Eq. \eqref{vangular} and follow the methodology outlined in \cite{ICO}. This yields $\Theta_{\rm GP}=\Omega-\Omega_{\rm GP}$, where $\Omega_{\text{GP}}$ is expressed as:
\begin{equation}
\Omega_{\rm GP}=\Omega\sqrt{A(r)}\sqrt{1-\frac{r^2\Omega^2}{A(r)}}\,.
\end{equation}
Using Eq. \eqref{dsKB} and expanding in small $l$ we get 
\begin{equation}
\Theta_{\rm GP}\approx \frac{3M^{3/2}}{2r^{5/2}}-\left(\frac{1}{2}+\frac{3M}{4r}\right)\sqrt{\frac{M}{r^3}}\,l \,.\label{tetafinal}
\end{equation}
The first term of Eq. \eqref{tetafinal} corresponds to the value obtained in GR, measured by the GP-B probe as $-6601.8\pm 18.3$ mas/yr. The second term represents the linear correction due to $l$. Utilizing data from the GP-B probe's gyroscope in Earth orbit, with mass $M_{\bigoplus}=4.4347\times 10^{-3}$ m and considering $r=(6371+642)\times 10^3$ m, we can constrain the parameter $l$ using Eq. \eqref{tetafinal}, yielding:
\begin{eqnarray}
-6.30714\times 10^{-12}\leq l\leq 3.90708\times 10^{-12}\,.
\end{eqnarray} 
These values are, in order of magnitude, very close to those obtained in \cite{KR} for the precession of Mercury. 

The geodesic precession angle of the probe changes accordingly \cite{presgeo},
\begin{equation}
\Delta\Phi=2\pi\left(1-\frac{\omega}{\Omega}\right)=2\pi\left(1-\frac{1}{\sqrt{A(r)-\frac{r}{2}A'(r)}}\right)\,.\label{deltageo}
\end{equation} 
Applying Eqs. \eqref{angularpropria} and \eqref{dsKB} in Eq. \eqref{deltageo} and expanding to powers of $\mathcal{O}\left(\frac{1}{r}\right)^2$ we get that
\begin{eqnarray}
\Delta\Phi\approx 2\pi\left(1-\sqrt{1-l}\right)-\Delta\Phi_{\rm GR}(1-l)^{3/2}\,,
\end{eqnarray}
where $\Delta\Phi_{\rm GR}=3\pi M/r= 1.22943$ mas, predicted for GR.  Using the measurements made by GP-B, we can measure the influence of the Lorentz symmetry breaking parameter on the change of direction of the spin of the gyroscope contained in the probe. Thus, for $l=-6.30714\times 10^{-12}$ we have
\begin{eqnarray}
\Delta\Phi\approx -1.229428675\,\text{mas}\,,
\end{eqnarray}
and for $l=3.90708\times 10^{-12}$
\begin{eqnarray}
\Delta\Phi\approx -1.229428674\,\text{mas}\,.
\end{eqnarray}
From these results, we can see that the effects of $l$ are very small and therefore difficult to detect on this scale.

\section{Circular  timelike Orbits}\label{time-like}

\subsection{ICO and ISCO orbits}

We are now interested in how the parameter of spontaneous Lorentz symmetry breaking can change the innermost circular orbit (ICO), representing the closest orbit a particle following a timelike geodesic can achieve without falling into the black hole. Orbits in this region are unstable and any perturbation will cause the particle to fall into the black hole.  Conversely, ICO differs from the innermost stable circular orbit (ISCO) in that orbits within ISCO are stable, allowing particles to orbit without being absorbed or expelled. However, particles orbiting within the ISCO gradually approach the black hole due to energy loss \cite{Shapiro}. 

Thus, we can infer that \cite{ICO}:
\begin{eqnarray}
C'(R)A(R)-C(R)A'(R)>0\,.
\end{eqnarray}
Using Eq. \eqref{dsKB} the above inequality takes the following form
\begin{eqnarray}
2\left(\frac{1}{1-l}-\frac{2M}{R}\right)-\frac{2M}{R}>0 \,,
\end{eqnarray} 
where
\begin{eqnarray}
3M(1-l)\leq R< 6M(1-l).
\label{Rico}
\end{eqnarray}
If $l=0$, we obtain the Schwarzschild result $R^{\rm Sch}_{\rm ICO}=3M$. The ICO radius is therefore modified by the spontaneous Lorentz symmetry breaking parameter, in that the $R_{\rm ICO}$ radius increases for $l<0$ and decreases for $0<l<1$.

\begin{figure*}[ht]
   \centering
      \includegraphics[scale=0.55]{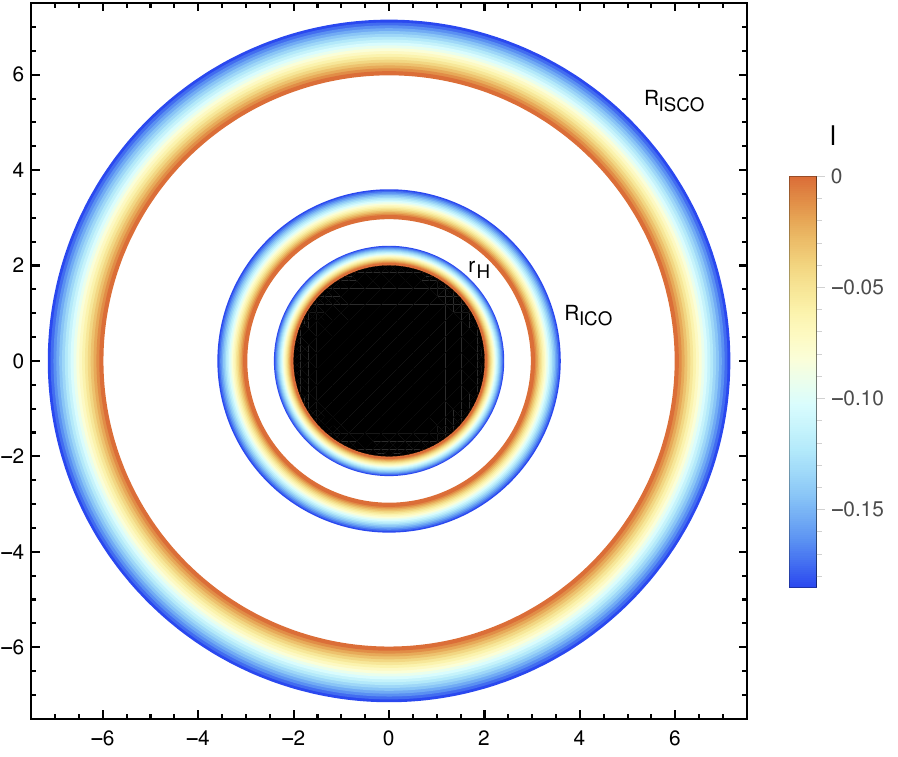}
      \includegraphics[scale=0.55]{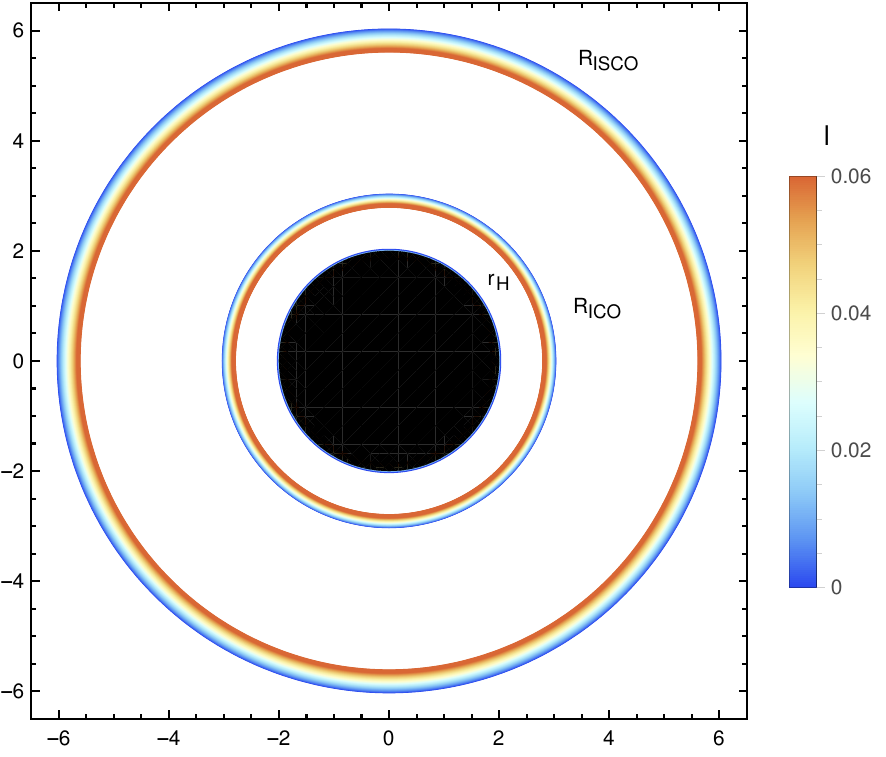}
    \caption{Representation of $R_{\rm ICO}$, $R_{\rm ISCO}$ and event horizon $r_H$ for values of $l<0$ and values of $0<l<1$.}
    \label{fig:RICOlneg}
\end{figure*}

Now we determine the ISCO radius. Consider $V''(r)=0\big|_{r=R_{\rm ISCO}}$, and from Eq. \eqref{Veff} we arrive at:
\begin{eqnarray}
M\left(\frac{6M}{r}-\frac{1}{1-l}\right)=0\Bigg|_{r=R_{\rm ISCO}} \,,
\end{eqnarray}
from which we deduce $R_{\rm ISCO}=6M(1-l)$, and obtain the ISCO radius for the Schwarzschild solution when $l=0$, i.e., $R^{\rm Sch}_{\rm ISCO}=6M$. Thus, $R_{\rm ISCO}$ increases slightly for values of $l<0$ and decreases slightly for values of $l>0$. This behavior can be seen in the parametric plot in Fig.\,\ref{fig:RICOlneg} and how the event horizon increases with negative $l$ values and decreases with positive $l$ values.

When the maximum and minimum of the potential merge we have two new conditions for $R_{\rm ISCO}$ to satisfy:
\begin{equation}
    V_{\rm eff}(r)=E^2\,, \qquad \ V'_{\rm eff}(r)=0.
\end{equation}
This allows us to obtain the energy $E_{\rm ISCO}$ and the momentum $L_{\rm ISCO}$ associated with ISCO and how $l$ changes their values. Thus, from the above conditions we have
\begin{eqnarray}
\frac{E_{\rm ISCO}}{M}=\frac{2}{3}\sqrt{\frac{2}{1-l}} \,,\label{Eisco}
\end{eqnarray} 
and 
\begin{eqnarray}
\frac{L_{\rm ISCO}}{M}=2\sqrt{3}\left(1-l\right)\,.\label{Lisco}
\end{eqnarray}
Note that for $l\rightarrow 0$ the values $E_{\rm ISCO}=\frac{2}{3}\sqrt{2}M$ and $L_{\rm ISCO}=2\sqrt{3}M$ are restored for the Schwarzschild geometry. The behavior of Eqs. \eqref{Eisco} and \eqref{Lisco} is plotted in Fig.\,\ref{Omb}, where we observe a subtle increase in energy and a decrease in angular momentum as $l$ increases.

\subsection{Zoom-whirl orbit}

\begin{figure}[t!]
      \includegraphics[scale=0.36]{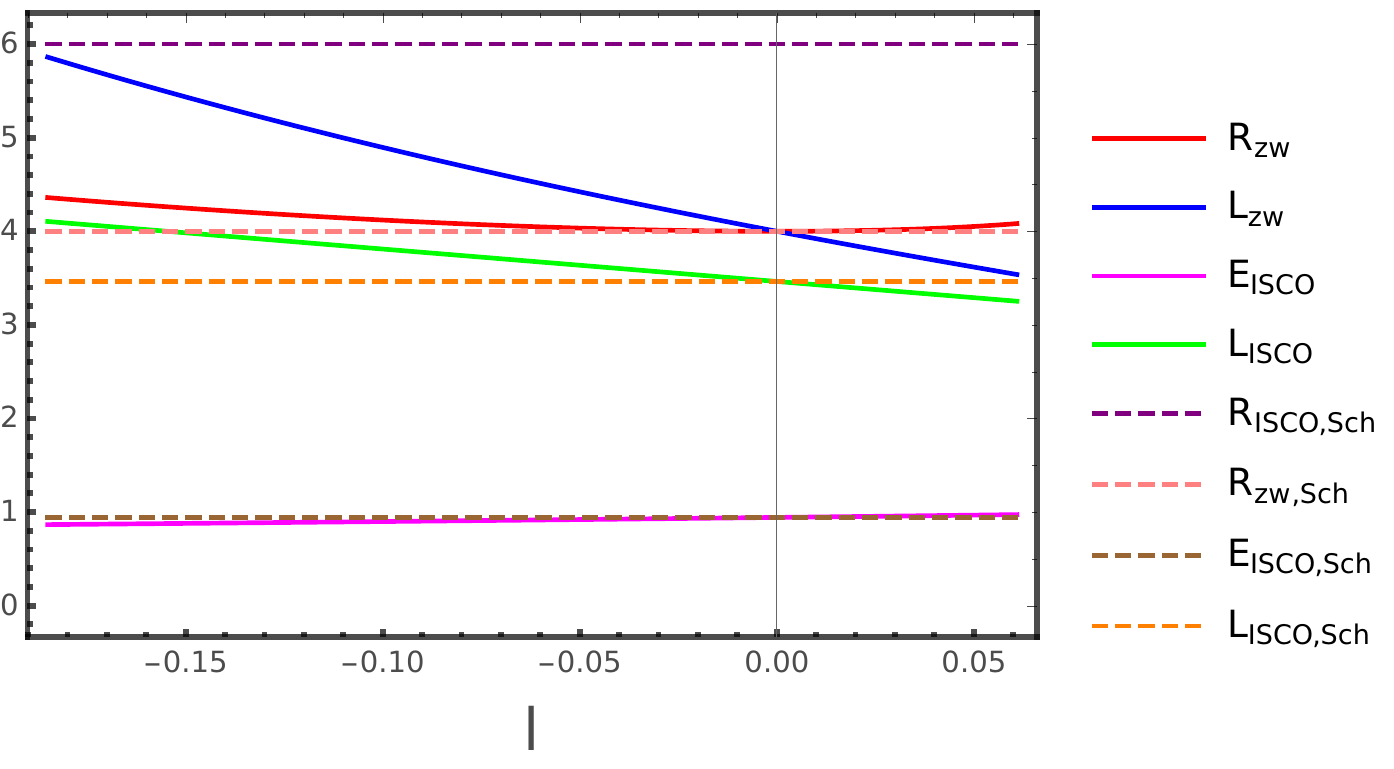}
    \caption{Behavior of the marginal orbit $R_{\rm zw}$ in red, the specific angular momentum $L_{\rm zw}$ in blue, the energy $E_{\rm ISCO}$ in cyan and the angular momentum $L_{\rm ISCO}$ in green, compared with values (dashed) for Schwarzschild. Here $R_{\rm zw, Sch}=L_{\rm zw, Sch}$ and all the quantities plotted here are divided by M.}
    \label{Omb}
\end{figure}

A particle obeying a timelike geodesic can still be trapped in an unstable circular orbit around the black hole, called a zoom-whirl orbit. This particle has zero binding energy and a specific angular momentum. These orbits, which are only marginally connected, are characterized by two main motions, the ``zoom'' motion, which is characterized by a rapid approach to the black hole, and the ``whirl'' motion, which is due to rotation around the black hole \cite{zoom1,zoom2}. To obtain the zoom-whirl orbit and the specific angular momentum and to check how it can be modified by $l$, we take the condition $V_{\rm eff}(r)=1$ in Eq. \eqref{Veff}, from which yields the angular momentum $L$:
\begin{eqnarray}
L=r\sqrt{\frac{1-A(r)}{A(r)}}\,.\label{Langular}
\end{eqnarray}
From the condition $V'_{\rm eff}=0$ and Eq. \eqref{dsKB} we obtain the radius of the zoom-whirl orbit, given by
\begin{eqnarray}
\frac{R_{\rm zw}}{M}=\frac{8}{4 l+\sqrt{1-8 l}+1}\,,\label{Rmb}
\end{eqnarray}
and using Eq. \eqref{Langular}, we obtain the angular momentum 
\begin{eqnarray}
\frac{L_{\rm zw}}{M}=\frac{8 \sqrt{\sqrt{1-8 l}+1}}{\sqrt{3-\sqrt{1-8 l}} \left(4 l+\sqrt{1-8 l}+1\right)}\,.\label{Lespecifico}
\end{eqnarray}
From this expression the limits for Schwarzschild solution are restored if $l\rightarrow 0$. The behavior of Eqs. \eqref{Rmb} and \eqref{Lespecifico} is plotted in Fig.\,\ref{Omb}, which shows that $R_{\rm zw}$ decreases for values of $l<0$ and increases slightly for values of $l>0$. $L_{\rm zw}$, on the other hand, decreases with increasing $l$.

\section{Circular lightlike orbit}\label{LIGHT-LIKE}

We now restrict the $l$ parameter of the solution \eqref{dsKB} to light-type geodesics, which satisfy $\mathcal{L}=0$. We are therefore interested in verifying how close this circular geodesic can be to the external event horizon $r_H$.

\subsection{Lower and upper limits of photon sphere}

\begin{figure*}[ht]
   \centering
      \includegraphics[scale=0.55]{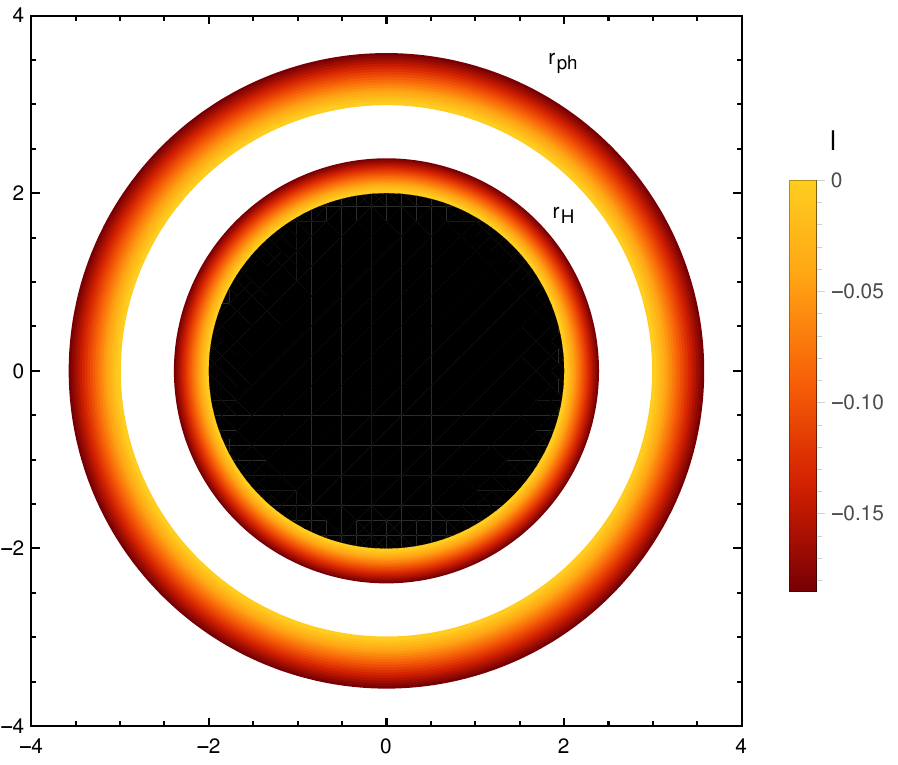}
      \includegraphics[scale=0.55]{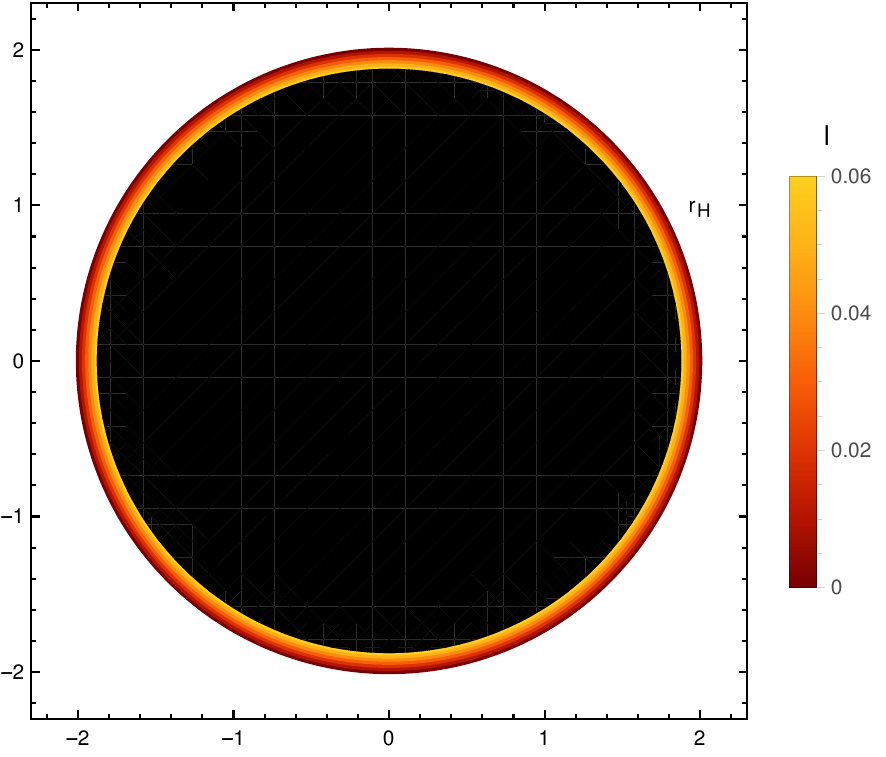}
    \caption{Illustration of the photon sphere $r_{\rm ph}$ and the horizon radius $r_{\rm H}$ for values of $l<0$ (left frame) and horizon radius for values $0<l<1$ (right frame). }
    \label{fig:rhrphlneg}
\end{figure*}

We adopt the approach outlined in \cite{liminf}, which extends the conjecture asserting that the photon sphere of black holes is constrained below by $r_{\rm ph}\geq 3r_{\rm H}/2$. This extension encompasses not only spherically symmetric hairy black holes but also spherically symmetric and static black holes in general, as demonstrated by:
\begin{eqnarray}
\left(r_{\rm in}-\frac{3}{2}r_{\rm H}\right)A(r)\geq 0\,,\label{rinf}
\end{eqnarray} 
where $r_{\rm in}$ is the innermost null circular geodesic obtained by
\begin{eqnarray}
\frac{C'(r)}{C(r)}=\frac{A'(r)}{A(r)}\,.\label{rhp}
\end{eqnarray}
From Eq. \eqref{dsKB} we find that $r_{\rm in}=r_{\rm ph}=3M(1-l)$ represents the closest radius to the horizon where light maintains its circular trajectory without deviation.
The upper bound for the photon sphere radius must satisfy $A(r_{\rm ph})\leq 1/3$ \cite{lsup}, so for \eqref{dsKB} we get
\begin{eqnarray}
\frac{1}{1-l}-\frac{2M}{r_{\rm ph}} \leq \frac{1}{3}\,.
\end{eqnarray}
This implies that $l\leq 0$ and $l>1$ for $M>0$, so that $l$ can take negative values. 
Although mathematically possible, values of $l>1$ are not allowed by definition because they change the signature of the metric to $(+,-,+,+)$, which means that $g_{00}$ never cancels and therefore there is no event horizon. The left frame of Fig.\,\ref{fig:rhrphlneg} shows how the radius of the photon sphere and the event horizon increase for negative values of $l$, and decrease in the event horizon for positive values of $l$, in the right frame.

\subsection{Event Horizon Telescope measurements}

We now constrain the parameter $l$ of the solution \eqref{dsKB} by the measurements of the EHT for Sgr A*.  For small angles, the shadow radius is given by \cite{EHT}
\begin{eqnarray}
r_{\rm sh}=r_{\rm ph}\sqrt{\frac{A(r_o)}{A(r_{\rm ph})}} \,,
\label{rsh}
\end{eqnarray}
which depends explicitly on the position $r_o$ of the static observer when it is close to the black hole. For a distance between the black hole and the observer at $r_o$ that is much larger than the gravitational radius then we have $A(r_0)\rightarrow 1/(1-l)$.

\begin{figure*}[ht]
   \centering
      \includegraphics[scale=0.4]{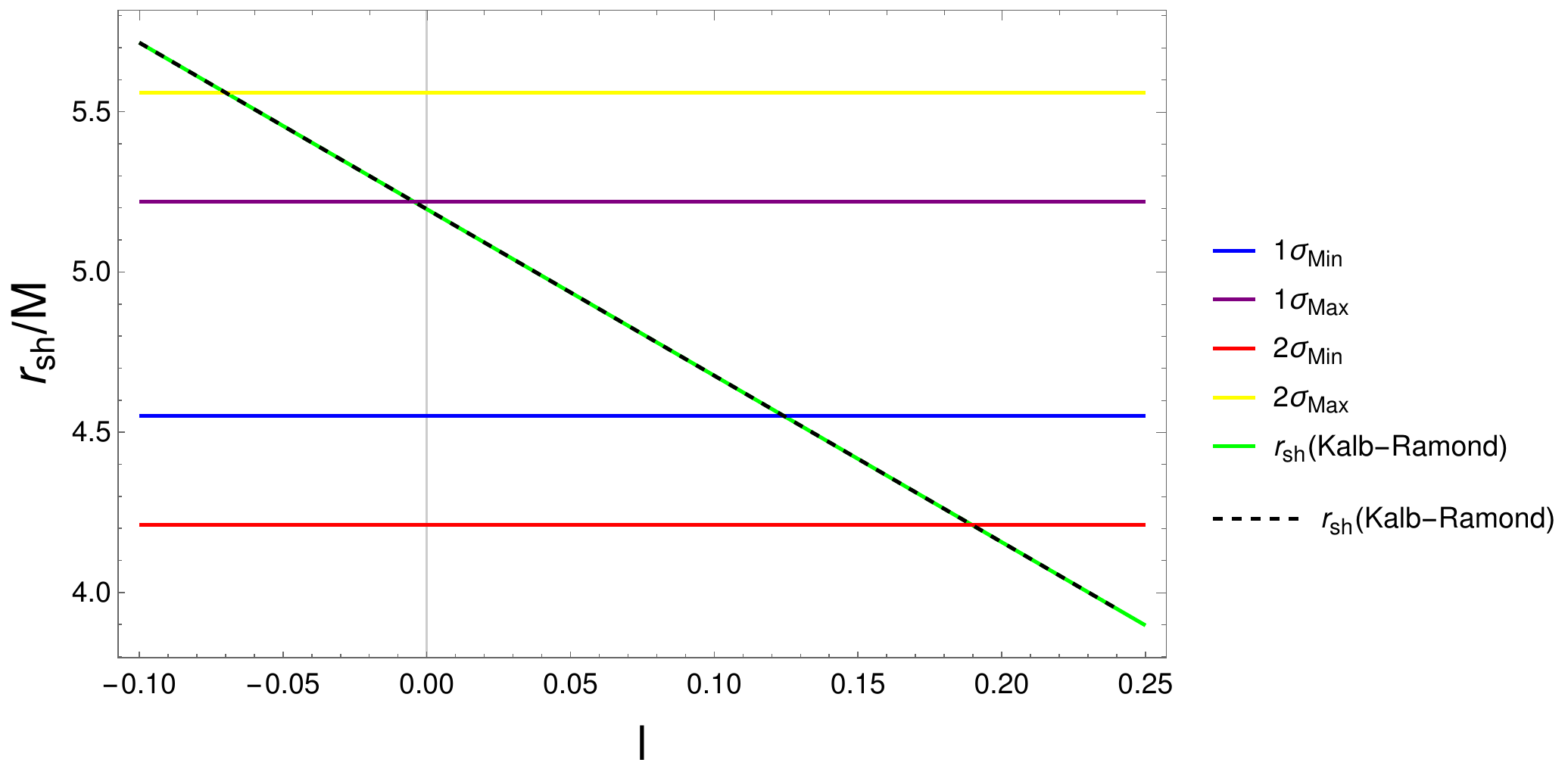}
    \caption{Graphical representation of $r_{\rm sh}$ as a function of $l$ for an observer at $r_o$ (in green) at a finite distance from the black hole and for an observer at infinity (dashed black).}
    \label{fig:rsp}
\end{figure*}

Let us now calculate the shadow radius $r_{\rm sh}$ for Eq. \eqref{dsKB} and compare it with the results measured by the EHT for the Sgr A* image \cite{EHT}. 
If we use the radius of the photon sphere $r_{\rm ph}=3M(1-l)$, we obtain in \eqref{rsh}
\begin{eqnarray}
r_{\rm sh}=3M(1-l) \sqrt{3+\frac{6M(1-l)}{r_o}}\,.\label{ro}
\end{eqnarray}
For $r_o\gg M$, the shadow radius is given by
\begin{eqnarray}
r_{\rm sh}\approx 3 \sqrt{3} M(1-l)\,.\label{rsh2}
\end{eqnarray}
Here it is easy to see that in the limit $l \rightarrow 0$ we recover the shadow radius for Schwarzschild solution, i.e.,  $r^{\rm Sch}_{\rm sh} = 3 \sqrt{3} M$. The measurements of the scaled image conducted by the EHT for GR place restrictions on  $r_{\rm sh}/M$, enabling us to impose a constraint on $l$ by comparing it with the theoretical value of $r_{\rm sh}$ in our case. In Fig.\,\ref{fig:rsp} (in green) we illustrate the behavior of the shadow radius, given by \eqref{ro}, for an observer at a finite distance from the black hole, and \eqref{rsh2} (in dotted black) for an observer at infinity, as a function of the parameter $l$. We observe a decrease in the shadow radius with increasing $l$ within the interval $-0.0700225\leq l\leq 0.189785$.

Note that there is no difference between the observers. This is because the last term in the square root of Eq. \eqref{ro} is of the order of $10^{-10}$ for an observer at a distance of $r_o=7.953$kpc. This behavior is expected and shown for a number of spherically symmetric solutions in \cite{EHT}. From Fig. \ref{fig:rsp}, we see that the observations of EHT for Sgr A*  have the lower and upper bounds $4.55\lesssim r_{\rm sh}/M\lesssim 5. 22$, represented by the range bounded by $1\sigma_{\rm min}$ and $1\sigma_{\rm max}$ respectively, and $4.21\lesssim r_{\rm sh}/M\lesssim 5.56$, represented by the range bounded by $2\sigma_{\rm min}$ and $2\sigma_{\rm max}$. These limits were determined by the relation between the mass and the distance of Sgr A* derived by the \textit{Keck} and \textit{Very Large Telescope Inteferometer-VLTI} instruments by observing the orbit of the star S2 around the galactic center. For more details on the values mentioned here, we refer the reader to \cite{EHT} and references there.in 

\section{Summary and  conclusion}\label{Sec:Conclusion}
 
In this work, we considered a spherically symmetric and static solution of the Schwarzschild-like metric incorporating the Kalb-Ramond field, modified by a spontaneous Lorentz symmetry breaking parameter \cite{KR}. We compared the theoretical outcomes of KR gravity with the Schwarzschild metric, considering experiments conducted within the framework of the GRAVITY collaboration \cite{GRAVITY1,GRAVITY4,GRAVITY5}, observations from the GP-B probe \cite{GPROBB, GPB}, and data obtained from the Event Horizon Telescope  \cite{EHT}.

In Section \ref{sec:two}, we constrained the symmetry-breaking parameter for particles following timelike geodesics to values within $-0.185022 \leq l \leq 0.060938$. This was achieved through the calculation of the relativistic orbital precession of the periastron of the star S2 around Sgr A*, compared to the result obtained for GR by the GRAVITY collaboration. Notably, when compared to the result for the precession of Mercury's orbit around the Sun from \cite{KR}, our findings exhibit a difference of approximately 10 to 11 decimal places. Our approach, based on methodologies presented in \cite{P1, P3, P2}, addresses subtleties such as metric approximations and differences in masses, which likely contributed to this disparity.
 
In Section \ref{Geodesic Precession}, using measurements of the geodesic drift rate by GP-B for a gyroscope within the probe, we determined a validity interval for the spontaneous Lorentz symmetry-breaking parameter: $-6.30714\times 10^{-12}\leq l\leq 3.90708\times 10^{-12}$. Additionally, we assessed the effects of $l$ on the gyroscope's spin direction change, concluding its negligible impact at this scale. Consequently, we infer that $l$ effects become more pronounced in intense gravitational fields, like those near black holes. Notably, the obtained range for $l$ aligns with other results in this study. An immediate extension could involve investigating gyroscope precession around a supermassive black hole, likely expanding the parameter's validity range.

We showed in Section \ref{time-like} that both the ICO and ISCO trajectories change simultaneously, either increasing or decreasing, while the values of $R$ consistently remain within the range defined by Eq.\,\eqref{Rico}. The parameter $l$ plays a crucial role, influencing the size of the black hole and consequently altering the ICO and ISCO orbits compared to the Schwarzschild case within the suggested $l$ intervals. We identified the zoom-whirl orbits $R_{\rm zw}$ lying between the ISCO and the marginally bound orbit for the Schwarzschild solution, $R_{\rm zw, Sch}=4M$, and observed that $R_{\rm zw}$ approaches $R_{\rm zw, Sch}$ for $l < 0$ and moves slightly away for $l > 0$. The most noticeable effect of spontaneous symmetry-breaking is observed in the behavior of the momentum $L_{\rm zw}$ and $L_{\rm ISCO}$, where there is a significant decrease as $l$ increases within its range.

In Section \ref{LIGHT-LIKE}, building upon the conjecture established in \cite{liminf} regarding spherically symmetric hairy black hole configurations, where the photon sphere radius is bounded below by $r_{\rm ph}\geq\frac{3}{2}r_{\rm H}$ and above by $A(r_{\rm ph})\leq1/3$ \cite{lsup}, we demonstrate that the parameter $l$ should only be confined to values $l\leq 0$. Values of $l>1$ are disregarded due to the metric signature change, implying the absence of an event horizon. Figure \ref{fig:rhrphlneg} illustrates the variations of the photon sphere and the event horizon with respect to the spontaneous Lorentz symmetry-breaking parameter. This constraint remains consistent with experimental tests and the $l$ values derived in Section \ref{sec:two} for Sgr A*.

Continuing within the domain of lightlike geodesics, we compared the shadow radius $r_{\rm sh}$ for two distinct observers: one situated at a finite distance from the KR black hole, and the other positioned at an infinite distance. These comparisons were made in accordance with the constraints established by the scale images generated by the EHT for the GR solution. Following this analysis, the determined range for $l$ is $-0.0700225\leq l\leq 0.189785$. Notably, there exists no discrepancy between the values obtained for each observer, which can be clearly seen in Fig.  \ref{fig:rsp}. Moreover, we demonstrate that within the permissible range, the shadow radius diminishes with increasing $l$. This same behavior exhibited by the KR metric due to spontaneous symmetry breaking was obtained for several spherically symmetric solutions in \cite{EHT}.
Experimentally, the $l$ parameter has been verified in the phenomenon of gravitational waves, as shown in 
\cite{Amarilo:2023wpn}.

The results presented in this paper align with experimental observations, particularly when the $l \rightarrow 0$ limit is considered. While we have derived stringent bounds on the possible $l$-values, it is important to note that these bounds do not conclusively determine the precise value of the spontaneous Lorentz symmetry-breaking parameter.
We have the possibility of measuring some observables in the gravitational lensing phenomenon, which depend on the $l$ parameter, thus showing its influence \cite{Junior:2024vdk}. Further validation of this parameter can be pursued through additional gravitational tests, thereby ensuring comprehensive scrutiny of its implications.

Examples of new constraints on the $l$ parameter can be obtained in the analysis of quasi-normal modes \cite{Cardoso:2008bp}, quasi-periodic oscillations from black hole \cite{Ingram:2019mna}, the concordance between the orbits of Sagittarius A$^{*}$ S-type stars \cite{s2conclu}, and finally, the analysis of black hole thermodynamics \cite{Rodrigues:2022qdp}. We trust that the constraints resulting from these analyses will complement those found within the present work.
In a new work, we are also checking the influence of the $l$ parameter on the memory of gravitational waves.

\section*{Acknowledgements}

MER thanks Conselho Nacional de Desenvolvimento Cient\'ifico e Tecnol\'ogico - CNPq, Brazil, for partial financial support. This study was financed in part by the Coordena\c{c}\~{a}o de Aperfei\c{c}oamento de Pessoal de N\'{i}vel Superior - Brasil (CAPES) - Finance Code 001.
FSNL acknowledges support from the Funda\c{c}\~{a}o para a Ci\^{e}ncia e a Tecnologia (FCT) Scientific Employment Stimulus contract with reference CEECINST/00032/2018, and funding through the research grants UIDB/04434/2020, UIDP/04434/2020 and PTDC/FIS-AST/0054/2021.
DRG is supported by the Spanish Agencia Estatal de Investigación Grant No. PID2022-138607NB-I00, funded by MCIN/AEI/10.13039/501100011033, FEDER, UE, and ERDF A way of making Europe.



\end{document}